# Impact of locus coeruleus and its projections on memory and aging

Jason Langley[1], Sana Hussain[2], Daniel E. Huddleston[3], Ilana J. Bennett[4], and Xiaoping P. Hu[1,2]*

1. Center for Advanced Neuroimaging, University of California Riverside, Riverside, CA
2. Department of Bioengineering, University of California Riverside, Riverside, CA
3. Department of Neurology, Emory University, Atlanta, GA
4. Department of Psychology, University of California Riverside, Riverside, CA

* Correspondence to:
Xiaoping P. Hu, Ph.D.
E-mail: xhu 'at' engr.ucr.edu

**Abstract**

*Introduction:* Locus coeruleus (LC) is the primary source of norepinephrine to the brain and its efferent projections innervate many brain regions, including the thalamus. LC degrades with normal aging, but not much is known regarding whether its structural connectivity evolves with age or predicts aspects of cognition.

*Methods:* Here, we use high-resolution diffusion tensor imaging (DTI)-based tractography to examine structural connectivity between LC and the thalamus in younger and older adults.

*Results:* We found LC projections to be bundled in a fiber tract anatomically consistent with the central tegmental tract (CTT) and branch from this tract into the thalamus. The older cohort exhibited a significant reduction in mean and radial diffusivity within CTT as compared to the young cohort. We also observed a significant correlation between CTT mean, axial, and radial diffusivities and memory performance (delayed recall) in the older adult cohort.

*Discussion:* These observations suggest that although LC projections degrade with age, the degree of degradation is associated with cognitive abilities in older adults.

**Keywords**
Locus coeruleus, DTI, aging, central tegmental tract, memory

**Significance Statement**
Locus coeruleus modulates several cognitive processes including modulates arousal, attention modulation, and memory. Sustaining the integrity of locus coeruleus neurons is hypothesized to play a key role in staving off age-related cognitive decline. However, less is known about how efferent projections of locus coeruleus change with age or cognition. Here, we examine how age effects microstructure of the central tegmental tract, a fiber tract in which locus coeruleus efferent projections are bundled, and whether age-related changes in the microstructure of this tract are associated with cognitive decline.

# 1. Introduction

Locus coeruleus (LC) is a small cylindrical structure located along the dorsal edge of the brainstem anterior to the 4th ventricle that is filled with catecholaminergic neurons. As the primary source of norepinephrine to much of the brain, it innervates many areas including the neocortex, thalamus, and hippocampus (Aston-Jones and Cohen, 2005). The release of norepinephrine by neurons in LC is hypothesized to reduce inflammation and protect neurons from oxidative stress (Feinstein, et al., 2002; Robertson, 2013; Troadec, et al., 2001), with the density of LC neurons correlating with cognitive performance in older adults (Kelly, et al., 2017; Wilson, et al., 2013). As such, LC is thought to play a pivotal role in forestalling cognitive decline, such as what occurs in normal aging.

Neuromelanin-sensitive contrast, associated with incidental (Sasaki, et al., 2006) or explicit (Chen, et al., 2014; Huddleston, et al., 2018; Schwarz, et al., 2011) magnetization transfer effects, has been found to colocalize with melanized neurons in LC (Keren, et al., 2009; Keren, et al., 2015). LC regions of interest (ROIs) derived from magnetization transfer contrast have been employed to examine how LC modulates attention (Krebs, et al., 2018; Murphy, et al., 2014), memory encoding (Clewett, et al., 2014; Clewett, et al., 2018; Hammerer, et al., 2018), or examine resting state functional connectivity between LC and memory networks (Jacobs, et al., 2015; Jacobs, et al., 2018). Other studies have found that structural measures within LC, such as magnetization transfer contrast (Clewett, et al., 2016; Dahl, et al., 2019) or diffusivity (Langley, et al., 2020b), correlate with memory performance in older adults. Together, these studies suggest that MRI can be used as a proxy for assessing norepinephrine function (Cassidy, et al., 2019) and its contribution to cognition in adults across the lifespan.

Histological studies have found a rostral-caudal gradient of age-related cell loss in LC, with rostral portions of LC exhibiting the greatest neuronal loss (Chan-Palay and Asan, 1989; Manaye, et al., 1995). This is consistent with imaging studies that have found reduced neuromelanin-sensitive contrast in rostral LC ROIs of older adults (Betts, et al., 2017; Dahl, et al., 2019; Liu, et al., 2018). Using retrograde tracers in macaques, noradrenergic fibers arising from the rostral portions of LC have been shown to enter the dorsomedial area of the central tegmental tract (CTT) and a group of fibers branch out to the dorsal thalamus. The remaining fibers in the noradrenergic bundle move adjacent to the ventral surface of the thalamus, merge into the medial forebrain bundle, and branch through the substantia innominata toward the amygdala (Bowden, et al., 1978; Tanaka, et al., 1982). Age-related degradation of the noradrenergic bundle may account for decreased catecholaminergic innervation observed in several brain regions by tracer studies in older primates (Goldman-Rakic and Brown, 1981). However, less is known about how LC projections in the noradrenergic bundle vary with age or relate to memory in humans.

Diffusion tensor imaging (DTI) tractography can be used to reconstruct fiber tracts *in vivo* (Mori and van Zijl, 2002) and their integrity can be examined using measures such as fractional anisotropy (FA), which gives an estimate of restricted diffusivity in the fiber bundle. In this study, we build on prior work examining age-related differences in LC microstructure and explore how age affects microstructure of the CTT using a high resolution DTI protocol. Our focus here is on the CCT given that this first segment of the noradrenergic bundle has defined endpoints (LC, thalamus) and is well established in histology. We further assess the relationship between white matter integrity of the CTT and memory performance. Finally, we investigate the rostral-caudal selectivity of age-related gray matter microstructural differences in LC and their effect on memory.

# 2. Materials and Methods

Sixty-six participants (30 older and 36 younger participants) were recruited for this study from the Riverside, California community and the student population at University of California, Riverside, respectively. All participants in the study gave written informed consent in accordance with local institutional review board (IRB) regulations. Prior to enrollment in the study, potential participants were screened over the phone for normal global cognition as defined by a score of >17 on a subset of the Montreal Cognitive Assessment (MOCA) adapted for phone screening (Pendlebury, et al., 2013). They were also screened for contraindications to MRI imaging and diagnosed neuropsychiatric condition, such as depression or stroke, that could influence their performance. One younger subject was excluded from the diffusion-weighted MRI analysis due to problems with data acquisition and three participants (2 older and 1 younger) were excluded from the analyses because of significant motion artifacts. The final sample size was 28 older and 35 younger participants. Demographic information (gender, age), MOCA, and Rey Auditory Verbal Learning Test (RAVLT) scores were collected on each subject. Group means for age, MOCA, and RAVLT scores are given in Table 1.

**Table 1.** Demographic data for the final sample are presented as mean ± standard deviation. MOCA - Montreal cognitive assessment scoring; RAVLT - Rey Auditory Verbal Learning Test.

| Variable | Younger (n=35) | Older (n=28) | $p$ Value |
|---|---|---|---|
| Gender (M/F) | 15/20 | 13/15 | 0.19 |
| Age (years) | 20.7±2.2 | 72.5±6.9 | $<10^{-4}$ |
| MOCA score | 27.5±1.6 | 27.1±1.7 | 0.35 |
| RAVLT Total | 49.8±7.5 | 41.7±12.5 | 0.003 |
| RAVLT Immediate | 11.5±2.2 | 7.6±3.6 | $<10^{-4}$ |
| RAVLT Delay | 11.1±2.5 | 7.8±3.7 | 0.0001 |

*2.1 Image Acquisition*

Imaging data were acquired on a 3 T MRI scanner (Prisma, Siemens Healthineers, Malvern, PA). Excitation was performed using the body coil on the scanner and signal was received using a 32-channel receive only coil at the Center for Advanced Neuroimaging at University of California, Riverside. Anatomic images were acquired with an MP-RAGE sequence (echo time (TE)/repetition time (TR)/inversion time=3.02/2600/800 ms, GRAPPA acceleration factor=2, flip angle=8°, voxel size=0.8×0.8×0.8 mm$^3$) for registration from subject space to common space. High-resolution diffusion-weighted MRI data were collected with a diffusion-weighted single-shot spin-echo, echo planar imaging sequence with the following parameters: TE / TR = 78 / 4500 ms, FOV = 194 × 168 mm$^2$, matrix size of 204 × 176, voxel size = 0.95 × 0.95 × 1 mm$^3$, multiband factor = 2, and 64 slices with no gap, covering the brain from the middle of the cerebellum to the striatum. Monopolar diffusion-encoding gradients were applied in 30 directions with $b$ values of 500 s/mm$^2$ and 2000 s/mm$^2$. Two sets of $b$=0 images were acquired, with the two sets having opposite polarities of phase-encoding direction for the correction of susceptibility distortion (Andersson, et al., 2003).

*2.2 Standard space transformation*

Imaging data were analyzed with FMRIB Software Library (FSL). A transformation was derived between individual subject space to Montreal Neurological Institute (MNI) 152 T$_1$-weighted space using FMRIB's Linear Image Registration Tool (FLIRT) and FMRIB's Nonlinear Image Registration Tool (FNIRT) in the FSL software package using the following steps (Smith, et al., 2004; Woolrich, et al., 2009). (1) The T$_1$-weighted image was skull stripped using the brain extraction tool (BET) in FSL, (2) brain extracted T$_1$-weighted images were aligned with the MNI brain extracted image using an affine transformation, and (3) a nonlinear transformation (FNIRT) was used to generate a transformation from individual T$_1$-weighted images to T$_1$-weighted MNI152 common space.

*2.3 Diffusion processing*

Diffusion-weighted data were analyzed with FSL (Jenkinson, et al., 2002; Jenkinson and Smith, 2001; Smith, et al., 2004) and MATLAB (The Mathworks, Natick, MA). Standard preprocessing steps were applied to correct susceptibility induced distortions in the diffusion MR data. Diffusion MR data were first corrected for eddy-current distortion, motion, and for susceptibility distortion using eddy in FSL (Andersson, et al., 2003; Andersson and Sotiropoulos, 2016). Next, skull stripping of the $T_1$-weighted image and susceptibility corrected $b=0$ image was performed using the brain extraction tool in the FSL software package (Smith, 2002). Measures derived from the diffusion MR data, including FA, mean diffusivity (MD), radial diffusivity (RD), and axial diffusivity (AD) were calculated using the dtifit tool in FSL. A bi-exponential model of water diffusion was used to estimate the fractional contribution of freely diffusing (i.e., unrestricted) water to the overall signal per voxel (Hoy, et al., 2014; Pasternak, et al., 2009). The free water elimination algorithm was implemented in DiPy (Garyfallidis, et al., 2014).

Diffusion-weighted data from each subject was assessed for gross head motion and ghosting artifacts from pulsatile motion. Three subjects were excluded from the analysis due to significant motion artifacts and had root-mean-square (RMS) translation estimate between consecutive volumes greater than 4 voxels. For the remaining subjects used in the analysis, the maximal RMS translation estimate between consecutive volumes was 1.53 voxels (standard deviation: 0.74 voxels).

Registration between diffusion-weighted and $T_1$ images was derived using the $b=0$ image with a rigid body transformation with boundary based registration cost function. The quality of each registration was assessed for each subject and no significant misregistrations were observed across the sample.

*2.4 LC tractography*

Prior to fiber tracking, principal diffusion directions and crossing fibers were estimated using dtifit and bedpostX. Next, brain extraction of the $b=0$ image was performed using BET (Smith, 2002). Segmentation of grey matter (GM), white matter (WM), cerebrospinal fluid and their partial volume estimates were created from the anatomical scans using FAST (Zhang, et al., 2001). Probabilistic tractography, as implemented in FSL, was used to track the fiber tracts originating from LC to thalamus.

A LC atlas was derived from magnetization transfer images in 31 healthy participants (mean age: 26.1 years) using processes similar to those described previously (Langley, et al., 2020b). This atlas was used to define a LC seed mask for tractography separately for each hemisphere. A bilateral thalamus atlas from the Harvard-Oxford atlas (http://fsl.fmrib.ox.ac.uk/fsl/fslwiki/atlases) was selected as the target mask. The procedures used to transform bilateral seed and target masks in diffusion space were as follows: Each mask was transformed to each subject's DTI-space using the nonlinear transformation described above concatenated with a rigid-body transformation using applywarp. The boundary between gray matter GM in the thalamus was found using the following procedure. First, the thalamus and WM partial volume estimates were transformed to DTI space. Next, the partial volume estimates of WM were thresholded at a level of 0.5, dilated, and binarized. Finally, the overlap of the thalamus mask and processed WM maps were defined as the thalamus and WM boundary mask.

For each subject, probabilistic tractography was carried out from all voxels in each seed mask for both hemispheres using a model accommodating two fiber orientations (Behrens, et al., 2007) in DTI space. Probability density functions were calculated by drawing 5000 samples per voxel in the LC seed mask from the uncertainty distribution of the fiber orientations between the LC seed mask and the thalamus and WM boundary mask. The number of these samples that reached the thalamus and WM target mask was counted as the waytotal. After each subject's LC-thalamus tracts were tracked, the pathways were normalized by the waytotal number, and the normalized value in each voxel was interpreted as the tract probability given the prior knowledge that the tract exists in that voxel. Then, each subject's tract probabilities were transformed to MNI space using a nonlinear transformation. Next, all transformed tracts in MNI space were averaged and multiplied by 100 to generate a percentile track probability for LC-thalamus tract in older and young populations. The process for creating the tract atlas is summarized in Figure 1. Finally, the young LC-thalamus tract was then transformed back to native diffusion space and mean FA, MD, AD, RD, and free water were measured in the resulting bilateral LC-thalamus population tract for each subject.

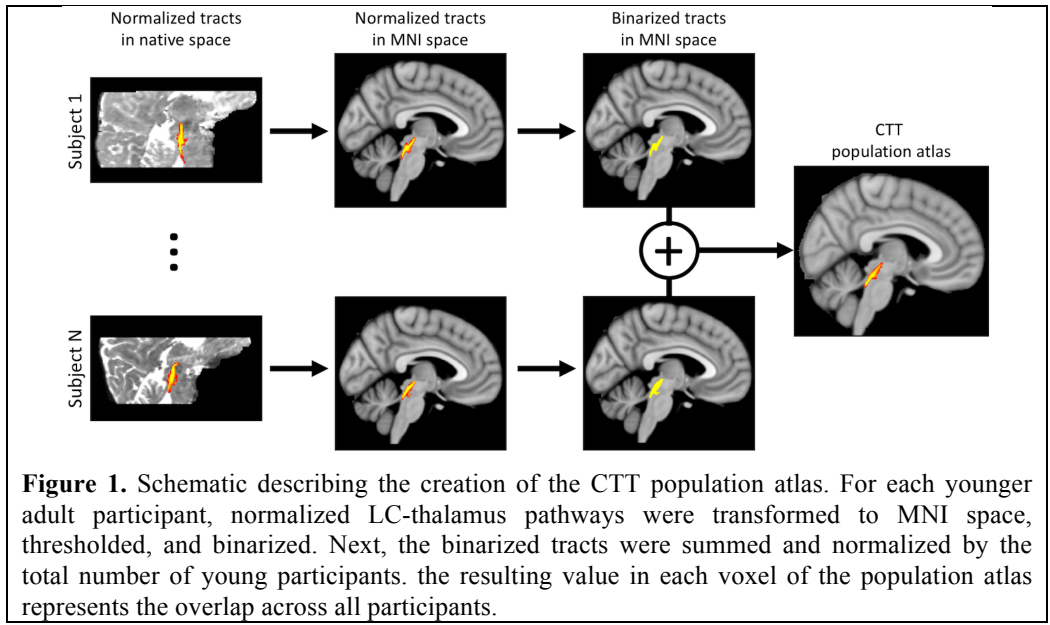

**Figure 1.** Schematic describing the creation of the CTT population atlas. For each younger adult participant, normalized LC-thalamus pathways were transformed to MNI space, thresholded, and binarized. Next, the binarized tracts were summed and normalized by the total number of young participants. the resulting value in each voxel of the population atlas represents the overlap across all participants.

*2.6 LC subregion analysis*

The LC atlas was divided into rostral and caudal ROIs with each ROI having a volume of 50% of the total volume of the LC atlas. Each ROI was transformed to diffusion space, thresholded, and binarized in an identical procedure to that of the LC seed mask described above. The mean volume of the bilateral rostral and caudal ROIs in diffusion space was each approximately 30 voxels. Mean FA, MD, RD, and AD were measured in caudal and rostral LC ROIs for each subject.

*2.7 Nigrostriatal tractography*

A nigrostriatal tract atlas was created to test the specificity of the relationship between CTT diffusivity and memory performance as well as diffusivity of LC subregions. A substantia nigra pars compacta atlas was derived from magnetization transfer images in 31 healthy participants (mean age: 26.1 years) using processes similar to those described previously (Langley, et al., 2020a; Langley, et al., 2017) was used to define the nigral seed mask. A bilateral putamen atlas from the Harvard-Oxford atlas (http://fsl.fmrib.ox.ac.uk/fsl/fslwiki/atlases) was selected as the target mask and tractography was performed in the younger adult group as described in section 2.6. The nigrostriatal tract atlas was transformed to native diffusion space, thresholded at a level of 0.6, and binarized. For each subject, mean FA, MD, RD, and AD were measured in the nigrostriatal atlas. Results for the nigrostriatal analysis are summarized in the Supplemental Materials section.

*2.8 Statistical analysis*

All statistical analyses were performed using IBM SPSS Statistics software version 24 (IBM Corporation, Somers, NY, USA) and results are reported as mean ± standard deviation. A *p* value of 0.05 was considered significant for all statistical tests performed in this work. Normality of diffusion data was assessed using the Shapiro-Wilk test for each group and all data was found to be normal.

We hypothesized that degradation of CTT, as measured by DTI indices, is correlated with cognitive measures in the older adult group but not in the younger adult group. Thus, to assess the impact of age-related decline in this tract, we performed a Spearman rank correlation of mean CTT DTI indices with cognitive measures (RAVLT delayed recall scores) separately in younger and older adult groups. Age group differences in the magnitude of these relationships were assessed using a one-tailed Fisher's z-test for independent correlations. We further hypothesized that the rostral LC subregion has stronger correlations with cognitive measures than the caudal LC ROI in older adults. This hypothesis was tested by performing a Spearman rank correlation of RAVLT delayed recall with rostral or caudal LC subregions separately in the older adult group. LC subregion ROI differences in the magnitude of these relationships were assessed using a one-tailed Fisher's z-test for independent correlations.

The effects of age group (younger, older) and LC subregion (caudal, rostral) were tested with separate two-way analyses of variance (ANOVA) for each diffusion index (MD, AD, RD). If the interaction was significant, *post hoc* comparisons between the age groups were performed for each LC subregion using respective two-tailed *t*-tests. Age group comparisons of the LC-thalamus tract (CTT) were made using separate two-tailed *t*-tests for each diffusion parameter.

Finally, to examine the relationship between CTT microstructure and microstructure of LC subregions, Spearman rank correlations were performed on each CTT DTI index to the corresponding LC subregion DTI index (i.e. CTT MD was correlated with rostral LC MD). LC subregion differences in the magnitude of these relationships were assessed using a one-tailed Fisher's z-test for independent correlations.

## 3. Results

*3.1 LC tractography*

The existence and variability of fiber tracts between LC and thalamus across individuals were assessed by measuring the maximum intensity value at any given voxel. The intensity value at each voxel in Figure 2 represents the amount of overlap across individuals for that voxel since the tracts for each individual were binarized and then summed to develop group maps for each tract. The maximum probability of the fiber tract in each group was 1. The variability of fiber tracts between LC and thalamus was further probed by examining the total volume of the tract with intensity values greater than 0.75. Approximately 53% of the tract volume had a probability value of 0.75 or greater in the younger group. Both results indicate significant overlap and consistency between individual tractography results. Importantly, tracts leaving the LC seed mask were contained within a region anatomically consistent with the CTT and projecting to the dorsal portion of the thalamus (see Figure 2).

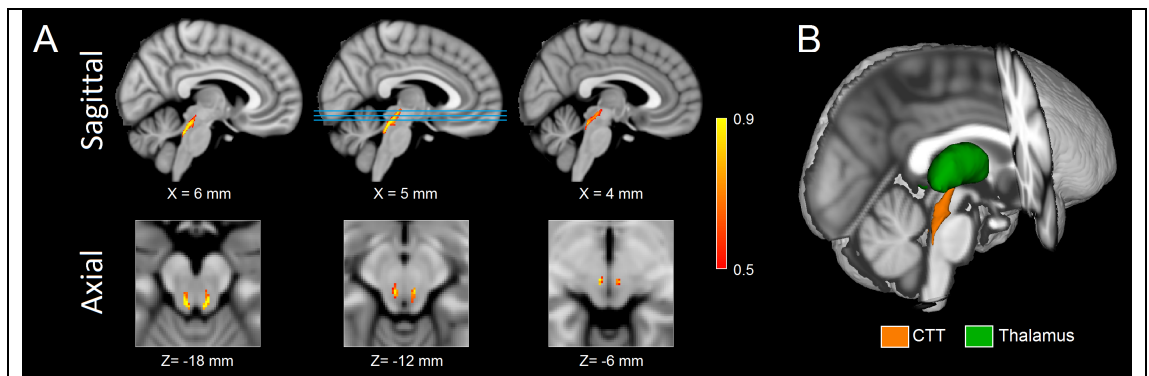

**Figure 2.** Sagittal (A; top row) and axial (A; bottom row) of population maps showing the tract from LC to thalamus in the young group. The lines in sagittal view at X=5 mm identify the location of the axial slices shown in the bottom row. The population map was generated by taking individual tracts, thresholding at a level of 0.15 (corresponding to 15% of the total streamlines sent from LC), binarizing, and transforming to standard space. The tract is anatomically consistent with the CTT. A three dimensional rendering of the CTT is shown in B.

*3.2 Age-related CTT integrity analyses*

Relative to young adults, higher FA (older: 0.31±0.03; young: 0.28±0.03; $t$=-2.6, $p$=0.006) and lower MD (older: $3.14\times10^{-4}$ mm$^2$/s ± $2.9\times10^{-5}$ mm$^2$/s; young: $3.31\times10^{-4}$ mm$^2$/s ± $3.2\times10^{-5}$ mm$^2$/s; $t$=2.4, $p$=0.009) were observed in the CTT in the older adult cohort. These differences were driven by lower RD within the tract in the older cohort (older: $3.14\times10^{-4}$ mm$^2$/s ± $2.9\times10^{-5}$ mm$^2$/s; young: $3.31\times10^{-4}$ mm$^2$/s ± $3.2\times10^{-5}$ mm$^2$/s; $t$=3.0, $p$=0.002). No difference in AD (older: $4.20\times10^{-4}$ mm$^2$/s ± $4.0\times10^{-5}$ mm$^2$/s; young: $4.31\times10^{-4}$ mm$^2$/s ± $3.8\times10^{-5}$ mm$^2$/s; $t$=1.3, $p$=0.197) or free water (older: 0.34 ± 0.05; young: 0.36 ± 0.04; $t$=1.43; $p$=0.158) was seen between cohorts. These results are summarized in Figure 3.

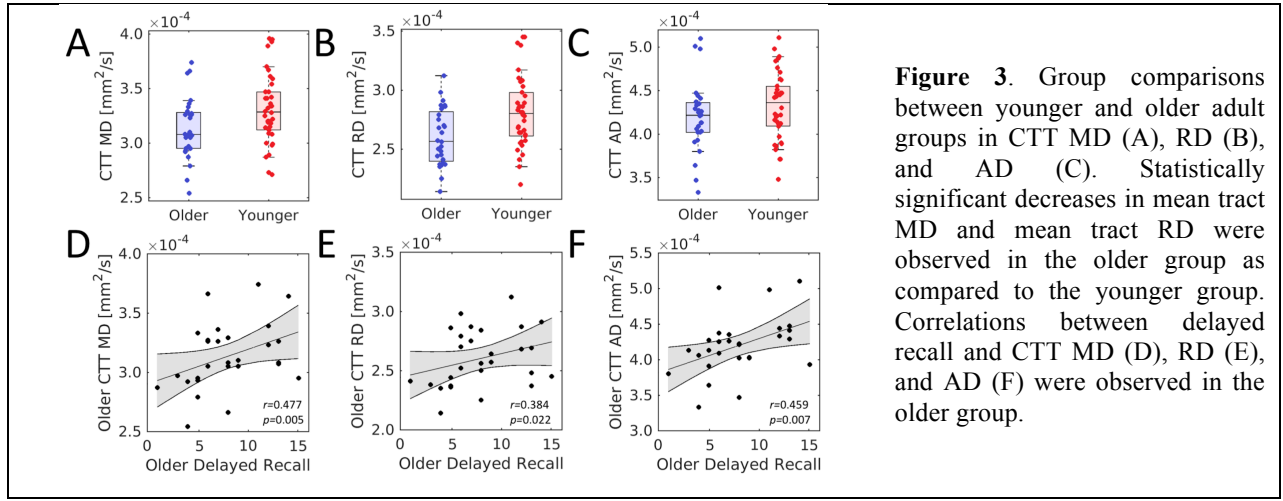

**Figure 3**. Group comparisons between younger and older adult groups in CTT MD (A), RD (B), and AD (C). Statistically significant decreases in mean tract MD and mean tract RD were observed in the older group as compared to the younger group. Correlations between delayed recall and CTT MD (D), RD (E), and AD (F) were observed in the older group.

*3.3 CTT integrity relates to memory performance.*

As shown in Figure 3, significant correlations were seen between diffusion metrics in the CTT and RAVLT delayed recall in the older group (mean tract MD: $r$=0.477, $p$=0.005; mean tract RD: $r$=0.384, $p$=0.022; mean tract AD: $r$=0.459, $p$=0.007) but not in the young group (mean tract MD: $r$=-0.268, $p$=0.104; mean tract RD: $r$=-0.309, $p$=0.06; mean tract AD: $r$=-0.173, $p$=0.282), with higher diffusivity relating to better memory performance. Fisher's $z$ tests revealed that these relationships were significantly weaker in the young compared to older group for all diffusion measures (mean tract MD: $z$=-2.79, $p$=0.003; mean tract RD: $z$=-2.71, $p$=0.007; mean tract AD: $z$=-2.51, $p$=0.012). No correlation was observed between CTT free water and RAVLT delayed recall in the older group ($r$=-0.098; $p$=0.347) but a significant correlation was observed between CTT free water and RAVLT delayed recall in the younger group ($r$=-0.392; $p$= 0.032). Controlling for age, education, and gender did not substantially change the correlations between CTT diffusion metrics and RAVLT delayed recall in the older adult group (mean tract MD: $r$=0.373, $p$=0.028; mean tract RD: $r$=0.296, $p$=0.046; mean tract AD: $r$=0.428, $p$=0.013) or in younger adult group (mean tract MD: $r$=-0.267, $p$=0.111; mean tract RD: $r$=-0.310, $p$=0.062; mean tract AD: $r$=-0.175, $p$=0.299).

*3.4 LC subregion analyses*

The effect of age and LC subregion on diffusivity was assessed using separate 2 Age Group x 2 LC subregion ANOVAs for each diffusion parameter. Results revealed significant main effects of LC subregion for all diffusion measures (MD: $F$=60.9, $p<10^{-4}$; RD: $F$=61.3, $p<10^{-4}$; AD: $F$=70.9, $p<10^{-4}$), with increased diffusivity in the rostral versus caudal LC subregion for MD (rostral: $3.89\times10^{-4}$ mm$^2$/s ± $4.4\times10^{-5}$ mm$^2$/s; caudal: $3.50\times10^{-4}$ mm$^2$/s ± $5.4\times10^{-5}$ mm$^2$/s), RD (rostral: $3.35\times10^{-4}$ mm$^2$/s ± $4.2\times10^{-5}$ mm$^2$/s; caudal: $2.98\times10^{-4}$ mm$^2$/s ± $5.1\times10^{-5}$ mm$^2$/s), and AD (rostral: $4.99\times10^{-4}$ mm$^2$/s ± $5.3\times10^{-5}$ mm$^2$/s; caudal: $4.51\times10^{-4}$ mm$^2$/s ± $6.0\times10^{-5}$ mm$^2$/s). Results for the comparison between rostral and caudal LC subregions are summarized in Figure 4. Main effects of age group were significant for RD ($F$=4.8, $p<0.032$), trending for MD ($F=3.5$; $p=0.066$), and not significant for AD ($F=0.001$; $p=0.98$), with increased diffusivity in younger versus older adults for RD (younger: $3.23\times10^{-4}$ mm$^2$/s ± $4.4\times10^{-5}$ mm$^2$/s; older: $3.03\times10^{-4}$ mm$^2$/s ± $5.0\times10^{-5}$ mm$^2$/s) and MD (younger: $3.76\times10^{-4}$ mm$^2$/s ± $4.7\times10^{-5}$ mm$^2$/s; older: $3.57\times10^{-4}$ mm$^2$/s ± $5.4\times10^{-5}$ mm$^2$/s). There were no significant interactions between age group and LC subregion for any diffusion measure (MD: $F=0.001$, $p=0.99$; RD: $F=0.102$, $p=0.689$; AD: $F=0.119$, $p=0.731$).

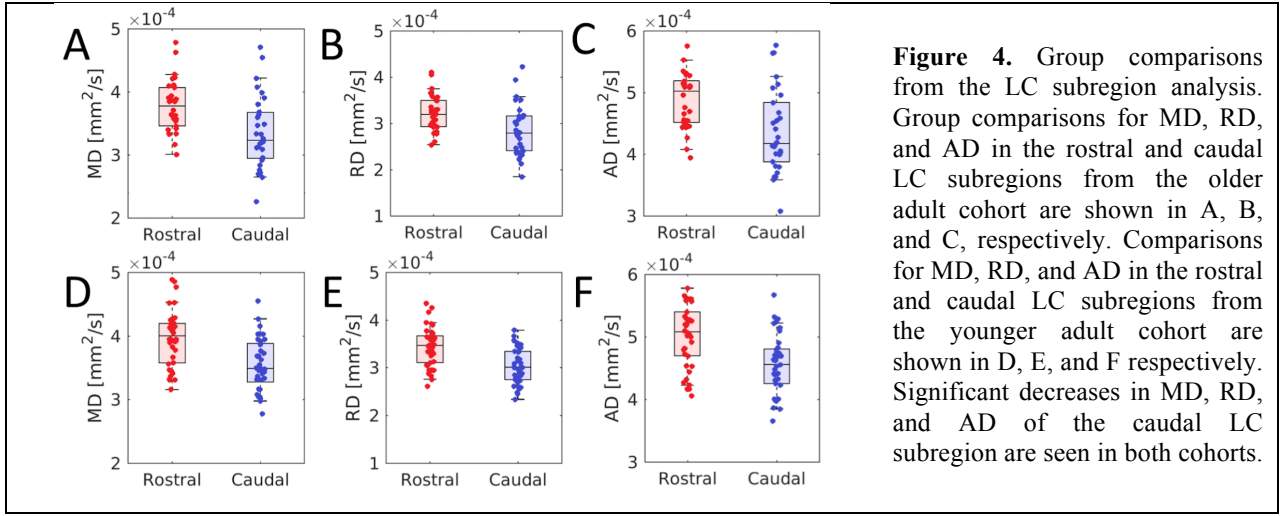

**Figure 4.** Group comparisons from the LC subregion analysis. Group comparisons for MD, RD, and AD in the rostral and caudal LC subregions from the older adult cohort are shown in A, B, and C, respectively. Comparisons for MD, RD, and AD in the rostral and caudal LC subregions from the younger adult cohort are shown in D, E, and F respectively. Significant decreases in MD, RD, and AD of the caudal LC subregion are seen in both cohorts.

3.5 *LC integrity relates to memory performance.*

For rostral LC, significant positive correlations were found between the RAVLT delayed recall score and MD ($r=0.645$; $p<10^{-4}$), RD ($r=0.614$; $p=1.6\times10^{-4}$), and AD ($r=0.626$; $p=1.1\times10^{-4}$) in the older group, with higher diffusivity being associated with better memory performance. In the young group, the RAVLT delayed recall score did not significantly correlate with MD ($r=-0.141$; $p=0.199$), RD ($r=-0.173$; $p=0.150$), or AD ($r=-0.117$; $p=0.242$). Fisher's *z* tests revealed that these relationships were significantly smaller in the young compared to older group for all diffusion measures (mean tract MD: $z=-3.42$, $p=0.0001$; mean tract RD: $z=-3.50$, $p=0.0005$; mean tract AD: $z=-3.19$, $p=0.0014$). These correlations for rostral LC are shown in Figure 5.

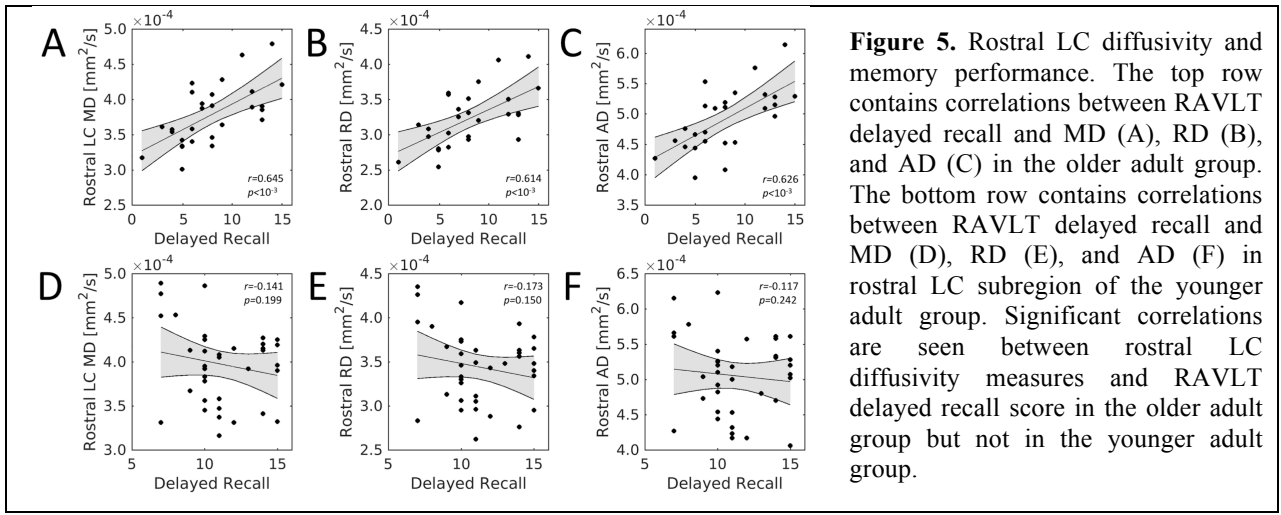

**Figure 5.** Rostral LC diffusivity and memory performance. The top row contains correlations between RAVLT delayed recall and MD (A), RD (B), and AD (C) in the older adult group. The bottom row contains correlations between RAVLT delayed recall and MD (D), RD (E), and AD (F) in rostral LC subregion of the younger adult group. Significant correlations are seen between rostral LC diffusivity measures and RAVLT delayed recall score in the older adult group but not in the younger adult group.

Langley, *et al.* LC Tractography 8

For caudal LC, significant positive correlations were also found between the RAVLT delayed recall score and MD ($r=0.606$; $p=1.9\times10^{-4}$), RD ($r=0.594$; $p=2.7\times10^{-4}$), and AD ($r=0.643$; $p<10^{-4}$) in the older group. However, no significant correlations were observed between the RAVLT delayed recall score and any diffusion measure (MD: $r=-0.009$, $p=0.478$; RD: $r=0.078$, $p=0.466$; AD: $r=-0.010$, $p=0.486$) in the younger group. Fisher's $z$ tests revealed that these relationships were significantly smaller in the young compared to older group for all diffusion measures (mean tract MD: $z=-2.67$, $p=0.0076$; mean tract RD: $z=-2.27$, $p=0.0232$; mean tract AD: $z=-2.51$, $p=0.0036$). These correlations for caudal LC are shown in Figure 6.

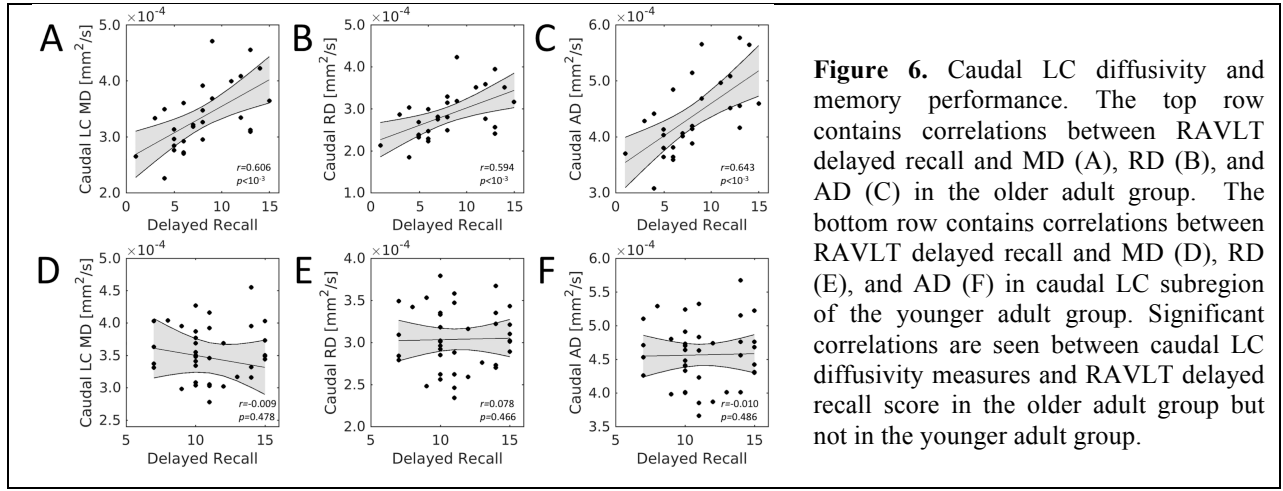

**Figure 6.** Caudal LC diffusivity and memory performance. The top row contains correlations between RAVLT delayed recall and MD (A), RD (B), and AD (C) in the older adult group. The bottom row contains correlations between RAVLT delayed recall and MD (D), RD (E), and AD (F) in caudal LC subregion of the younger adult group. Significant correlations are seen between caudal LC diffusivity measures and RAVLT delayed recall score in the older adult group but not in the younger adult group.

*3.6 LC subregion diffusivity relates to CTT diffusivity.*

For diffusivity indices in rostral LC, significant positive correlations were found between diffusivity indices in CTT (MD: $r=0.656$, $p=1.5\times10^{-4}$; RD: $r=0.507$, $p=0.006$; AD: $r=0.679$, $p=7.1\times10^{-5}$) in the older group, with higher diffusivity in rostral LC being associated with higher diffusivity in CTT. However, no correlation was observed between caudal LC MD and CTT MD ($r=0.292$, $p=0.132$), caudal LC RD and CTT RD ($r=0.165$, $p=0.400$), or caudal LC AD and CTT AD ($r=0.444$, $p=0.018$), after Bonferroni correction for 3 comparisons, in the older adult group. In the younger group, diffusion indices were highly correlated with CTT diffusion indices in both the rostral (MD: $r=0.787$, $p=4.7\times10^{-9}$; RD: $r=0.669$, $p=6.1\times10^{-10}$; AD: $r=0.822$, $p=1.5\times10^{-17}$) and caudal (MD: $r=0.492$, $p=0.002$; RD: $r=0.351$, $p=0.004$; AD: $r=0.561$, $p=8.1\times10^{-7}$) LC subregions. Fisher's $z$ tests revealed that these relationships were significantly smaller in the caudal LC compared to the rostral LC for all diffusivities in the older (MD: $z=-2.7$, $p=0.006$; RD: $z=-2.3$, $p=0.020$; AD: $z=-2.0$, $p=0.049$) and younger group (MD: $z=-2.7$, $p=0.007$; RD: $z=-2.4$, $p=0.023$; AD: $z=-2.7$, $p=0.009$). Correlations for the older cohort are shown in Figure 7.

## 4. Discussion

In furthering our current understanding of the role LC plays in cognition and age-related cognitive decline, it would be advantageous to map the white matter fiber tracts connecting LC with the rest of the brain. To this end, we focused on mapping LC efferent projections in the CTT, first segment of the noradrenergic bundle, since this connection has been well mapped by invasive tracer studies of non-human primates (Bowden, et al., 1978; Samuels and Szabadi, 2008; Tanaka, et al., 1982). After careful inspection of Duvernoy's Atlas of the Human Brain Stem and Cerebellum (Naidich, et al., 2009), we found LC efferent projections to be bundled in a fiber tract anatomically consistent with the CTT and branched from this tract into the thalamus. Age-related reductions in RD and increases in FA were observed in this tract and we found significant correlations between CTT diffusivity measures and RAVLT delayed recall in older, but not younger, adults. A complementary analysis of LC subregions found significant age-related microstructural differences in the rostral and caudal portions of LC (reduced MD and RD in older adults), with microstructure in both subregions correlating with RAVLT delayed recall in older, but not younger, adults. For both age groups, higher LC diffusivity was associated with higher CTT diffusivity in rostral, compared to caudal, subregions. Of note, none of these effects were observed in a control tract, the NST (see Supplemental Materials). Taken together, these findings suggest that axial connectivity and axonal size in CTT white matter play a role in the retention of cognitive abilities in aging.

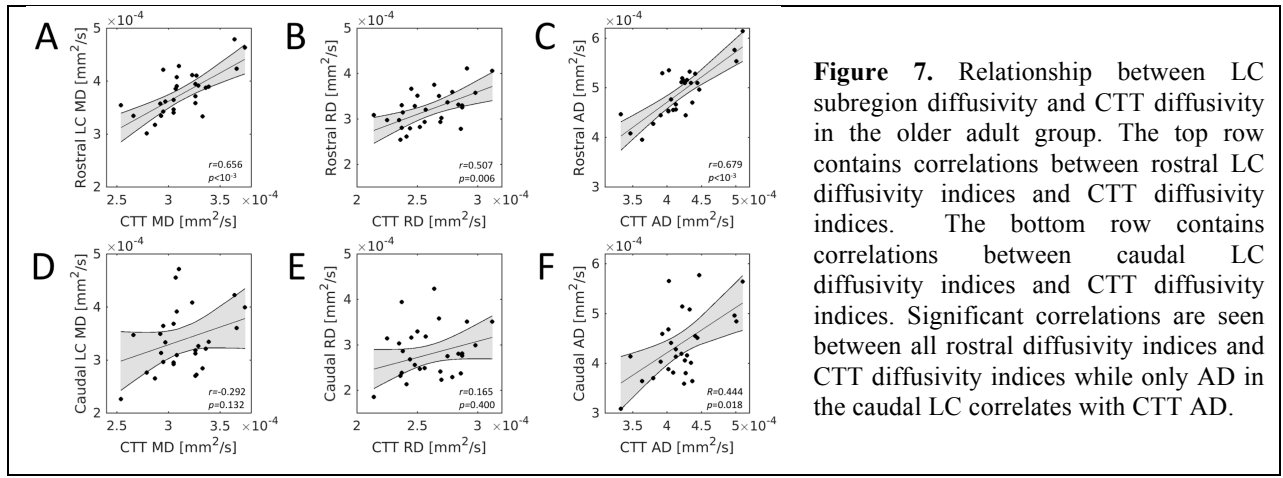

**Figure 7.** Relationship between LC subregion diffusivity and CTT diffusivity in the older adult group. The top row contains correlations between rostral LC diffusivity indices and CTT diffusivity indices. The bottom row contains correlations between caudal LC diffusivity indices and CTT diffusivity indices. Significant correlations are seen between all rostral diffusivity indices and CTT diffusivity indices while only AD in the caudal LC correlates with CTT AD.

White matter fiber tracts in CTT were identified by tractography from a seed region in LC to target region in the thalamus. Individuals demonstrated a high degree of overlap between tracts connecting LC and thalamus in standard space as indicated by the maximum overlap intensity and percent of tract volume shared by at least 75% of individuals in the younger adult group. This overlap measure is more accurately seen as a probability of tract existence. In particular, this type of probability measure reflects the anatomical variability that occurs across individuals and should not be thought of as a measure of connectivity strength (Hua, et al., 2009). In addition to finding that LC efferent projections are bundled in the CTT (Bowden, et al., 1978; Tanaka, et al., 1982), we observed higher FA and lower RD in this tract in the older adult group.

Efferent projections of LC are spatially organized with axons in the rostral portion projecting to the thalamus, hypothalamus, cortex, and forebrain while axons in the caudal portion project into the cerebellum and spinal cord (Bowden, et al., 1978; Tanaka, et al., 1982; Westlund, et al., 1984). Our analysis examining the relationship between CTT microstructural integrity and microstructural integrity of LC subregions found CTT microstructural integrity more closely related to microstructural integrity in the rostral LC subregion than in the caudal LC subregion in both cohorts. In particular, in the older adult cohort, which was expected to experience some age-related neuronal loss, we found significant correlations between CTT diffusivity and diffusivity in the rostral LC subregion but no relationship was observed between diffusivity in the caudal LC subregion and CTT diffusivity. These results accord with the aforementioned tracer studies (Bowden, et al., 1978; Tanaka, et al., 1982; Westlund, et al., 1984) and indicate that CTT diffusivity may be a proxy for LC integrity when scan resolution prohibits accurate LC segmentation in older adults.

The shape and density of cells changes along the rostral-caudal axis of LC. Cells are less dense and circularly shaped in the rostral portion of LC while the shape becomes less circular and density of cells increases in the caudal portions of LC (Chan-Palay and Asan, 1989; Manaye, et al., 1995). High neuronal density in the caudal LC should affect LC microstructure by increasing barriers to diffusion in that region (Beaulieu, 2002), which is expected to lead to decreases in diffusivity. Interestingly, rostral and caudal portions of LC in both cohorts had different microstructural properties with the caudal LC ROI exhibiting decreased diffusivity as compared to the rostral LC ROI. This result may reflect the differences in density of cells along the rostral-caudal axis of LC.

Age-related microstructural differences were observed in both rostral and caudal subregions of LC, with older adults showing reduced mean and radial diffusivity compared to younger adults. In contrast to our predictions, however, these age effects were not regionally selective. This observation is not in agreement with earlier neuromelanin-sensitive imaging studies which found an age-related reduction in neuromelanin sensitive contrast in the rostral portions of LC (Betts, et al., 2017; Dahl, et al., 2019; Liu, et al., 2018). The lack of regional selectivity with diffusion measures is not in accordance with several histology studies that found the rostral LC to exhibit extensive age-related neuronal loss (Chan-Palay and Asan, 1989; Manaye, et al., 1995), suggesting that these diffusivity measures may not be sensitive enough to detect subtle neuronal loss in this region. However, results from histology are not in complete agreement with several studies finding no age-related neuronal loss in the rostral portions of LC (Mouton et al., 1994; Ohm et al., 1997). These studies have speculated that the neuronal loss seen in other histology may be due to Alzheimer's disease since the rostral portion of LC is selectively affected by Alzheimer's disease (German et al., 1992; Hoogendijk et al., 1995).

Significant correlations were observed between reduced AD and RD in this tract and worse performance on RAVLT delayed recall in the older adult group. Reduced RD may indicate a winnowing of axons in CTT (Barazany, et al., 2009; Wheeler-Kingshott and Cercignani, 2009). We further found that lower performance on RAVLT delayed recall scores in the older adult group was correlated with reduced diffusivity in both rostral and caudal LC subregion ROIs. Interestingly, the correlations between RAVLT delayed recall and diffusivity in rostral and caudal LC were of similar strength and may suggest that there is no selectivity in participation of LC neurons in memory. This interpretation accords with both histology (Kelly, et al., 2017; Wilson, et al., 2013) and imaging (Clewett, et al., 2016) in older adults which found a reduction in total neuron count or contrast from melanized neurons in LC correlates with decreased cognitive performance.

In contrast, only a correlation between mean CTT free water and RAVLT delayed recall was seen in the younger cohort, with lower RAVLT delayed recall scores associated with higher CTT free water. Clinical studies assessing free water in white matter tracts have linked the free water measure to cellular density, atrophy, or neuroinflammation (Pasternak, et al., 2014; Pasternak, et al., 2012). Lower performance on RAVLT delayed recall is likely driven by individual differences in density of CTT as it is unlikely that there is significant atrophy or neuroinflammation in the CTT of a population of young control subjects.

No correlation was observed between RAVLT delayed recall and caudal or rostral LC diffusivity measures in the younger group. This result stands in contrast to a recently published imaging study that found higher neuromelanin-sensitive contrast in medial and caudal LC ROIs to be correlated with improved performance on an emotional memory task in younger adults (Jacobs, et al., 2020). RAVLT delayed recall is sensitive measure for diagnosis of cognitive impairment in older adults (Zhao, et al., 2015). However, younger adults, on average, perform better on RAVLT delayed recall than older adults (Messinis, et al., 2016) and younger adults tend to have similar performance on RAVLT delayed recall (Badcock, et al., 2011). Our data accords with these studies and the lack of correlation between diffusivity in LC subregions with RAVLT delayed recall may be due to the inability of RAVLT delayed recall to detect subtle differences in normal cognition in younger adults.

There are some caveats in the present study. First, neuromelanin-sensitive data were not acquired in the same cohort so it is impossible to examine microstructural changes in rostral and caudal LC subregion ROIs alongside contrast from melanized neurons. Second, we only considered fiber tracts verified by anatomical tracer studies in non-human primate and limited our tractography analysis to the CTT. Referencing anatomical tracer studies in non-human primates is necessary to interpret tractography results given the complexity and structural proximity of multiple pathways in the brainstem and ensure that no false tracts were identified. Finally, it is impossible to exclude the possibility that some of the healthy controls in the older adult cohort have an undiagnosed neurological condition which may affect the microstructure of LC or CTT.

## 5. Conclusion

In this work, a high-resolution diffusion-weighted MRI protocol was employed to examine age-related microstructural and compositional differences in LC subregions as well as in CTT. Older age was most associated with axonal thinning in rostral portions of LC and CTT. Strong correlations were observed between diffusivity and RAVLT delayed recall score in CTT and both subregions of LC. In particular, reduced diffusivity was correlated with worse memory performance in older adults and may suggest that connectivity and axonal size in white matter between LC and thalamus play a role in the retention of cognitive abilities.

## Author Contributions

JL contributed to the conception and design of the study, data analysis, conducted statistical analyses, and drafted the manuscript. SH contributed in data and statistical analyses. DH contributed to the interpretation of data. IJB contributed to the conception and design of the study, conducted statistical analyses, interpretation of data, and acquired financial support for the study. XH contributed to the conception, design of the study, and interpretation of data. All authors contributed to revisions of the manuscript.

## Funding

This work was supported by R00 AG047334 (Bennett) and R21 AG054804 (Bennett) from the National Institutes of Health/National Institute on Aging, 1K23NS105944-01A1 (Huddleston) from the National Institutes of Health/National Institute of Neurological Diseases and Stroke, and the American Parkinson's Disease Foundation Center for Advanced Research at Emory University (Huddleston).

## Disclosures

The authors have no conflicts of interest to disclose

**Supplementary Material**

*S1. Introduction*

A nigrostriatal tract (NST) atlas was created using procedure similar to that described in section 2.7 and used as a control tract to examine the specificity of central tegmental tract (CTT) diffusivity on memory performance and locus coeruleus (LC) subregion diffusivity.

*S2. Statistical Analysis*

All statistical analyses were performed using IBM SPSS Statistics software version 24 (IBM Corporation, Somers, NY, USA) and results are reported as mean ± standard deviation. A *p* value of 0.05 was considered significant for all statistical tests performed in this work. Normality of diffusion data was assessed using the Shapiro-Wilk test for each group and all data was found to be normal.

Unpaired t-tests were used to assess age group differences in each DTI index (FA, MD, AD, RD) from the NST. We performed separate Spearman rank correlations between mean NST DTI indices and memory performance (RAVLT delayed recall scores) separately for younger and older adults. Separate Spearman rank correlations were also performed between each NST DTI index and the corresponding DTI index from each LC subregion (rostral, caudal; i.e. NST MD was correlated with rostral LC MD) for each age group.

*S3. Results*

The older adult cohort exhibited higher FA (older: 0.28±0.03; young: 0.26±0.02; $t$=-2.6, $p$=0.007) and lower MD (older: $3.03\times10^{-4}$ mm$^2$/s ± $3.7\times10^{-5}$ mm$^2$/s; young: $3.45\times10^{-4}$ mm$^2$/s ± $4.2\times10^{-5}$ mm$^2$/s; $t$=4.2, $p$<0.001), lower RD (older: $2.65\times10^{-4}$ mm$^2$/s ± $3.3\times10^{-5}$ mm$^2$/s; young: $3.01\times10^{-4}$ mm$^2$/s ± $3.8\times10^{-5}$ mm$^2$/s; $t$=3.9, $p$<0.001), and lower AD (older: $3.82\times10^{-4}$ mm$^2$/s ± $4.4\times10^{-5}$ mm$^2$/s; young: $4.44\times10^{-4}$ mm$^2$/s ± $4.9\times10^{-5}$ mm$^2$/s; $t$=4.6, $p$<0.001) relative to young adults.

To probe the potential contribution of motion to the significant CTT-memory correlations observed within older adults, we tested for similar correlations between NST diffusion metrics and RAVLT delayed recall . No significant correlations were observed between diffusion metrics in NST and RAVLT delayed recall in the older group (mean tract MD: $r$=0.209, $p$=0.276; mean tract RD: $r$=0.165, $p$=0.392; mean tract AD: $r$=0.266, $p$=0.162) or in the younger group (mean tract MD: $r$=0.279, $p$=0.094; mean tract RD: $r$=0.280, $p$=0.094; mean tract AD: $r$=0.269, $p$=0.107). These results suggest that correlations between CTT diffusivity and memory in older adults are not due to motion. Group comparisons and correlations between NST and memory in the older adult group are summarized in Figure S1.

The specificity of the relationship between CTT diffusivity and LC subregion diffusivity was assessed by correlating NST and LC subregion diffusivity. No significant correlations were observed between rostral LC diffusivity indices and NST diffusivity (MD: $r$=0.209, $p$=0.276; RD: $r$=0.165, $p$=0.392; AD: $r$=0.266, $p$=0.162) in the older group or in the younger group (MD: $r$=0.046, $p$=0.787; RD: $r$=-0.001, $p$=0.957; AD: $r$=0.139, $p$=0.367). For the caudal LC ROI, no significant correlation was seen between caudal LC diffusivity indices and NST diffusivity (MD: $r$=0.195, $p$=0.247; RD: $r$=0.135, $p$=0.424; AD: $r$=0.247, $p$=0.140) in the younger group or in the older group (MD: $r$=0.111, $p$=0.938; RD: $r$=-0.089, $p$=0.666; AD: $r$=0.183, $p$=0.360). These results suggest that the correlation between CTT diffusivity and LC subregion diffusivity is specific to CTT.

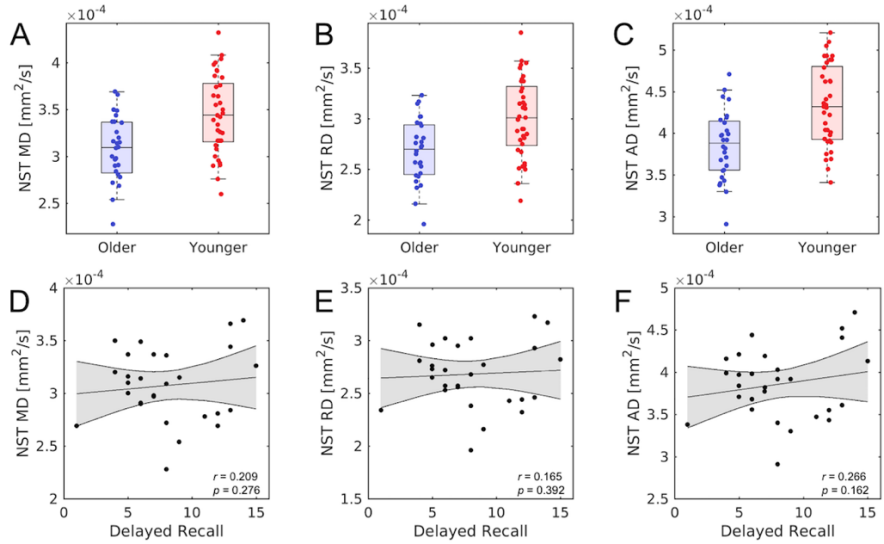

Figure S1. Group comparisons between younger and older adult groups in nigrostriatal (NST) MD (A), RD (B), and AD (C). Statistically significant decreases in mean tract MD, RD and AD were observed in the older group as compared to the younger group. No correlations between delayed recall and NST MD (D), RD (E), and AD (F) were observed in the older group.

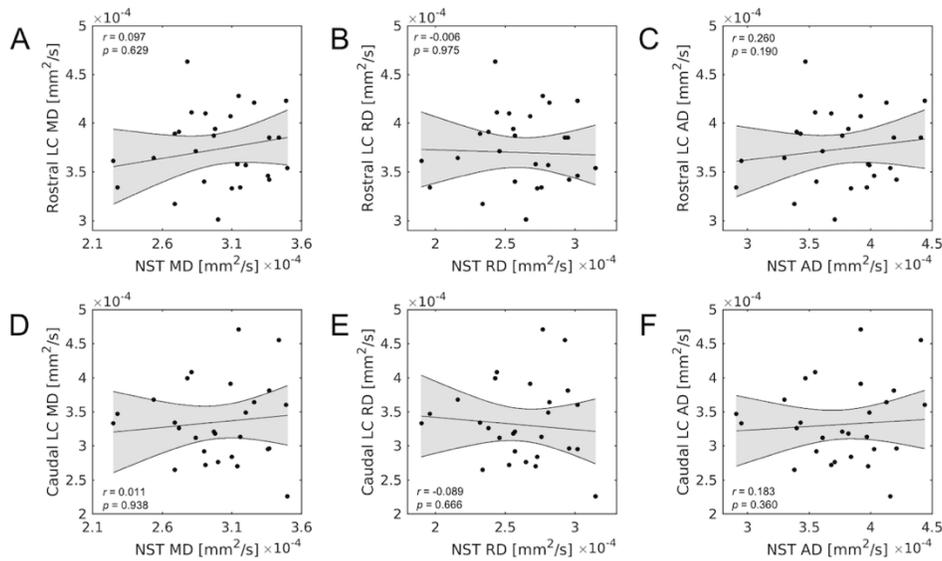

Figure S2. Relationship between LC subregion diffusivity and NST diffusivity in the older adult group. The top row contains correlations between rostral LC diffusivity indices and NST diffusivity indices. The bottom row contains correlations between caudal LC diffusivity indices and NST diffusivity indices. No significant correlations are seen between NST diffusivity indices and rostral or caudal LC diffusivity indices.